\documentclass[fleqn,twoside]{article}
\usepackage{espcrc2}

\usepackage{graphicx}
\usepackage[figuresright]{rotating}

\newcommand{\AmS}{{\protect\the\textfont2
  A\kern-.1667em\lower.5ex\hbox{M}\kern-.125emS}}

\title{Theoretical Expectations for Rare and Forbidden Tau Decays}

\author{Ernest Ma\address{Physics Department, University of 
        California, 
        Riverside, CA 92521, USA}}
       
\begin{document}

\begin{abstract}
Given the experimental evidence for $\nu_\mu - \nu_\tau$ oscillations, the 
existence of lepton flavor violation in $\tau$ decays is a theoretical 
certainty.  In this brief review, I consider the connection between models 
of neutrino mass and the expected observability of some $\tau$ decays.
\vspace{1pc}
\end{abstract}

\maketitle

\section{INTRODUCTION}

In the minimal standard model (SM) with $m_\nu = 0$, the 3 lepton numbers 
$L_e$, $L_\mu$, and $L_\tau$ are separately conserved.  However, given 
the present neutrino-oscillation data, which imply that neutrinos have 
mass and mix with one another, the only conserved lepton number is 
$L = L_e + L_\mu + L_\tau$ if $m_\nu$ is Dirac, and $(-1)^L$ if 
$m_\nu$ is Majorana.  As a result, lepton flavor violation must occur 
in $\tau$ decays, which may be classified into 3 groups:\\

\noindent (I) $\Delta L = 0$, and to be specific $\Delta L_\tau = -1$.\\

$\Delta L_\mu = 1$, $\Delta L_e = 0$ ~:~ $\tau^- \to \mu^- X$,\\

$\Delta L_\mu = 0$, $\Delta L_e = 1$ ~:~ $\tau^- \to e^- X$,\\

$\Delta L_\mu = 2$, $\Delta L_e = -1$ ~:~ $\tau^- \to \mu^- \mu^- 
e^+$,\\

$\Delta L_\mu = -1$, $\Delta L_e = 2$ ~:~ $\tau^- \to e^- e^- 
\mu^+$,\\

\noindent where $X = \gamma, \mu^-\mu^+, e^-e^+, q \bar q$.\\

\noindent (II) $\Delta L = \pm 1$.\\

$\tau^- \to n \pi^-$ ~[$\Delta L=-1, \Delta B=1$],\\

$\tau^+ \to p \pi^0$ ~[$\Delta L=+1, \Delta B=1$].\\

\noindent (III) $\Delta L = \pm 2$ ~[$\Delta (-1)^L = +$].\\

$\tau^- \to \mu^+ (e^+) d \bar u d \bar u$.\\

Group (I) is rare, but hopefully observable.  Group (II) is forbidden, to 
the extent that baryon number is conserved.  Group (III) is somewhere in 
between.

\section{MINIMAL EXPECTATIONS}

The present data on atmospheric and solar neutrino oscillations imply that, 
to a very good approximation, the neutrino mixing matrix is given by
\begin{equation}
\pmatrix {\nu_e \cr \nu_\mu \cr \nu_\tau} \simeq \pmatrix {c & -s & 0 \cr 
s/\sqrt 2 & c/\sqrt 2 & -1/\sqrt 2 \cr s/\sqrt 2 & c/\sqrt 2 & 1/\sqrt 2} 
\pmatrix {\nu_1 \cr \nu_2 \cr \nu_3}
\end{equation}
What does that tell us about $\tau$ decay?  Since lepton flavor is 
violated, $\tau \to \mu X$ and $\tau \to eX$ must exist, but they are not 
necessarily observable.\\

In analogy, consider the well-known $b \to s \gamma$ decay in the quark 
sector.  This proceeds in one-loop order through the exchange of a $W$ 
boson and $t,c,u$ quarks.  The reason that it is observable is twofold: 
(1) $V_{tb} \simeq 1$, $V_{ts} \simeq 0.04$, and (2) $m_t^2/M_W^2 \simeq 4.7$. 
On the other hand, the $\tau \to \mu \gamma$ decay involves neutrino 
intermediate states.  Although the mixing angle is maximum, i.e. $V_{\tau 3} 
= 1/\sqrt 2$, $V_{\mu 3} = -1/\sqrt 2$, the ratio $m_3^2/M_W^2$ is at best 
$10^{-21}$ which renders it totally unobservable.\\

The above argument applies if $m_\nu$ is Dirac, but if it is Majorana, 
then there is another contribution.  To see this, consider the famous 
seesaw neutrino mass matrix:
\begin{equation}
{\cal M}_\nu = \pmatrix {0 & m_D \cr m_D^T & M}.
\end{equation}
Let it be diagonalized by the unitary matrix:
\begin{equation}
U = \pmatrix {A & B \cr C & D},
\end{equation}
where $A A^\dagger + B B^\dagger = 1$.  Now $B$ is not zero but of order 
$m_D/M$, so the matrix $A$ which was used to connect $\tau$ and $\mu$ to 
the light neutrinos is not quite unitary.  The heavy neutrinos must also 
be considered as intermediate states.  Theoretically, $M > 10^{13}$ GeV is 
usually assumed, in which case $\tau \to \mu \gamma$ is again negligible. 
However, if $M$ is allowed to be much smaller, then there is some hope. 
Recently, it has been shown \cite{cvetic} that with certain assumptions of 
fine tuning and for $M < 10$ TeV, the largest $B(\tau \to \mu \gamma)$ is 
$10^{-9}$ and the largest $B(\tau \to 3 \mu)$ is $10^{-10}$, whereas the 
present experimental upper bounds are $1.1 \times 10^{-6}$ and $1.9 \times 
10^{-6}$ respectively. [New (preliminary) limits from BELLE reported at this 
Workshop are $6.0 \times 10^{-7}$ and $3.8 \times 10^{-7}$.]\\

Another recent analysis \cite{black} considers the effective operator 
$(\bar \mu \Gamma \tau)(\bar q^\alpha \Gamma q^\beta)$.  Using the various 
experimental upper bounds on $B(\tau \to \mu$ + hadrons), the scale of 
new physics is contrained up to about 10 TeV.

\section{INDUCED FLAVOR VIOLATION}

To have observable lepton flavor violation, there must be new physics at or 
below the TeV scale.  How is that possible if the scale associated with 
neutrino mass is very high?  The answer is supersymmetry where slepton 
flavor violation at the TeV scale can induce $\tau \to \mu (e)$ transitions. 
I will discuss two examples.

\subsection{Universal Soft Terms at the Planck Scale}

Consider the minimal supersymmetric standard model (MSSM) with the addition 
of 3 heavy singlet neutrino superfields ($N^c$).  The seesaw mechanism works 
as in the SM and the particle content of this model at low energies is the 
same as that of the MSSM.  There is however an important difference, namely 
the slepton mass matrix.  The reason is as follows.  Because of the existence 
of $N^c$, there is a Yukawa coupling matrix $Y_\nu$ which links $L = (\nu,e)$ 
to $N^c$.  Since neutrinos mix with one another, this matrix is not diagonal. 
This means that from the Planck scale ($M_P$) to $M$, the slepton mass 
matrix will pick up off-diagonal entries even though it starts out as 
universal, i.e. the supergravity scenario.  Specifically,
\begin{eqnarray}
(\Delta m^2_{\tilde L})_{ij} \simeq -{\ln (M_P/M) \over 16 \pi^2}  
[6m_0^2 (Y^\dagger_\nu Y_\nu)_{ij} && \nonumber \\ + 2(A^\dagger_\nu 
A_\nu)_{ij}], &&
\end{eqnarray}
where $i \neq j$, and $A_\nu$ is the soft supersymmetric breaking trilinear 
scalar coupling matrix.  The assumed boundary condition at $M_P$ is of course 
$m^2_{\tilde L} = m_0^2$.\\

To make contact with the actual neutrino mass 
matrix, the $M$ matrix itself has to be known, the simplest assumption 
being $M_1=M_2=M_3$.  If it is also assumed that $A_\nu$ is proportional 
to $Y_\nu$, then the above equation may be evaluated to find $m^2_{\tilde L}$ 
at the TeV scale.  The off-diagonal entries enable $\tau$ to decay into $\mu$ 
or $e$ in one-loop order through the exchange of sleptons and gauginos, 
which have masses at or below the TeV scale.  Many studies have been made 
regarding this possibility.  Typically \cite{ellis}, for a large class of 
neutrino textures,
\begin{equation}
B(\tau \to \mu \gamma) < 10^{-8},
\end{equation}
although $10^{-7}$ is possible at the edge of the allowed parameter space.\\

If the boundary conditions at $M_P$ are not enforced, there will not be any 
prediction.  In that case, it is possible \cite{kim} to have $B(\tau \to 
\mu \gamma)$ up to the present experimental bound, without 
any problem phenomenologically.\\

Going back to the case with universal boundary conditions, it has also been 
shown recently \cite{bk} that the neutral Higgs particles of the MSSM obtain 
off-diagonal couplings, thus allowing
\begin{equation}
B(\tau \to 3\mu) \simeq 10^{-7} \left( {\tan \beta \over 60} \right)^6 
\left( {100 ~{\rm GeV} \over m_A} \right)^4.
\end{equation}
Similarly, $B(\tau \to \mu \eta) \simeq 8.4 B(\tau \to 3\mu)$ is also 
possible \cite{sher}.

Note added:  After this talk was given, a preprint appeared \cite{der} 
which pointed out that in the above scenario, the corresponding $\tau \to \mu 
\gamma$ branching fraction would be much above the present experimental 
bound, thus $B(\tau \to 3 \mu) < 4 \times 10^{-10}$ is expected, rather 
than Eq.~(6).

\subsection{Radiatively Corrected Neutrino Mass Matrix}

Suppose that at some high scale, when the charged lepton matrix is 
diagonalized, the Majorana neutrino mass matrix is of the form
\begin{equation}
{\cal M}_\nu = \pmatrix {m_0 & 0 & 0 \cr 0 & 0 & m_0 \cr 0 & m_0 & 0}.
\end{equation}
Consider then the most general one-loop radiative corrections to the above. 
It is easily shown \cite{bmv} that ${\cal M}_\nu$ becomes
\begin{equation}
m_0 \pmatrix {1+2\delta+2\delta' & \delta'' & \delta''^* \cr 
\delta'' & \delta & 1+\delta \cr \delta''^* & 1+\delta & \delta},
\end{equation}
where only $\delta''$ is complex.  This matrix is very special, because in 
the limit $\delta''$ is real, it is exactly diagonalized by Eq.~(1).  In other 
words, the right ${\cal M}_\nu$ has been obtained almost without trying.\\

In the context of supersymmetry, $\delta$ comes from $\mu-\tau$ mixing 
in the slepton mass matrix, but unlike the previous example, it is not 
induced by the corresponding Yukawa matrix between $M_P$ and $M$.  In 
other words, the origin of soft supersymmetry breaking is not specified 
in this case; it is simply taken to be whatever is allowed by phenomenology. 
Let the two mass eigenvalues of the sleptons be $\tilde m_{1,2}$ and their 
mixing angle be $\theta$, then in the approximation $\tilde m_1^2 >> \mu^2 
>> M_{1,2}^2 = \tilde m_2^2$, where $\mu$ is the Higgsino mass and $M_{1,2}$ 
are the gaugino masses, the parameter $\delta$ is given by
\begin{eqnarray}
\delta &=& {\sin \theta \cos \theta \over 16 \pi^2} \left[ (3g_2^2-g_1^2) \ln 
{\tilde m_1^2 \over \mu^2} \right. \nonumber \\ && - {1 \over 4} 
\left. (3g_2^2+g_1^2) \left( \ln {\tilde m_1^2 \over \tilde m_2^2} - 
{1 \over 2} \right) \right].
\end{eqnarray}
Using $\Delta m^2_{32} \simeq 2.5 \times 10^{-3}$ eV$^2$ from atmospheric 
neutrino data, the condition
\begin{eqnarray}
&& \ln (\tilde m_1^2 / \mu^2) - 0.3 [ \ln (\tilde m_1^2 / \tilde 
m_2^2) - 1/2] \nonumber \\  && \simeq (0.535 / \sin \theta \cos \theta) 
(0.4 ~{\rm eV} / m_0)^2
\end{eqnarray}
is obtained.  This is easily satisfied if $m_0$ (the common mass measured 
in neutrinoless double beta decay) is not much less than the present 
experimental bound of about 0.4 eV.\\

Using the same approximation which leads to Eq.~(9), the $\tau \to \mu 
\gamma$ amplitude is then given by
\begin{equation}
{e(3g_2^2+g_1^2) \over 3072 \pi^2} \sin 2 \theta {m_\tau \over 
\tilde m_2^2} \epsilon^\lambda q^\nu \bar \mu \sigma_{\lambda \nu} (1+
\gamma_5) \tau.
\end{equation}
Using the experimental upper bound of $1.1 \times 10^{-6}$ ($6.0 \times 
10^{-7}$), this means that $\tilde m_2 > 102$ (119) GeV.\\

Any contribution to $\tau \to \mu \gamma$ must also contribute to the muon 
anomalous magnetic moment (but not vice versa).  Here, this contribution is 
less than about $10^{-10}$, which is well below the present experimental 
accuracy.\\

\section{DIRECT FLAVOR VIOLATION}

The new physics scale responsible for the neutrino mass matrix may be only a 
TeV. This is possible in many models, even in the seesaw case with 
right-handed neutrinos \cite{mr}.  Among such models, the most direct 
connection of the neutrino mass matrix to possible new physics is realized 
in the triplet Higgs model.  The interaction Lagrangian contains
\begin{equation}
f_{ij} \left[ \xi^0 \nu_i \nu_j + \xi^+ \left( {\nu_i l_j + l_i \nu_j \over 
\sqrt 2} \right) + \xi^{++} l_i l_j \right]
\end{equation}
which implies $({\cal M}_\nu)_{ij} = 2 f_{ij} \langle \xi^0 \rangle$.  It has 
recently been shown \cite{mrs} that very small $\langle \xi^0 \rangle$ may be 
obtained naturally, even if $m_\xi$ is only of the order 1 TeV.  In that case, 
the decay $\tau^- \to l^+_i l^-_j l^-_k$ may be observable through $\xi^{++}$ 
exchange.  Let the neutrino mass matrix be given by Eq.~(7), with 
$f_{\mu \tau} = f_{ee} = 0.12$ and $m_\xi = 1$ TeV, then
\begin{equation}
B(\tau^- \to \mu^+ e^- e^-) \simeq 1.3 \times 10^{-7},
\end{equation}
which is below the present experimental bound of $1.5 \times 
10^{-6}$, and close to the new (preliminary) limit of $2.8 \times 10^{-7}$ 
from BELLE.  Note that if $\xi^{++}$ could be produced at future colliders, 
its decay into same-sign charged leptons would map out the neutrino mass 
matrix, up to an overall scale.

\section{GROUP III DECAYS}

These decays change $L$ by two units, so they are allowed if $m_\nu$ is 
Majorana.  However, if the only $\Delta L = \pm 2$ terms in the Lagrangian 
come from $m_\nu$, then neutrinoless double beta decay is the only viable 
experimental signature.  If not, then $\tau^- \to \mu^+ d \bar u d \bar u$ 
may have a chance, but it would require something very exotic, such as 
a heavy neutral Majorana fermion $X$ with four-fermion interactions of the 
form $(\bar u d)(\bar l_R X)$.

\section{CONCLUSION}

Flavor violation in $\tau$ decays must occur at some level.  There are 
models of neutrino mass (whether the scale of new physics is $10^{13}$ GeV 
or 1 TeV), which predict such decays within reach of future experiments 
for a reasonable range of parameters.

\section*{Afterword}

This talk was given on September 11, 2002.  One year ago, my former Ph.D. 
student and colleague D. Ng was on the 80th floor of the North Tower when 
the first plane hit.  He was one of the last people who got out safely 
before it collapsed.

\newpage
\section*{Acknowledgement}

I thank Abe Seiden and the other organizers of the 7th International 
Workshop on Tau Lepton Physics for their great hospitality at Santa Cruz. 
This work was supported in part by the U.~S.~Department of Energy under 
Grant No.~DE-FG03-94ER40837.


\end{document}